# NURTURING LANGUAGE PROFICIENCY IN SPANISH-SPEAKING CHILDREN THROUGH DIGITAL COMPETENCE

**Rhayza Adela Jolley Rangel**

*Rey Juan Carlos University (SPAIN)*

## Abstract

This article explores the development of a digital platform designed to transform early-age English education, particularly for Spanish-speaking children. Focused on innovative methodologies, engaging visuals, and a comprehensive phonics approach, the platform emphasizes usability, accessibility, and user-centred design. English phonics, crucial for cognitive development, is taught from the age of two, addressing pronunciation challenges. The platform integrates adapted activities, gamification, and visual codes for effective learning. This initiative stands as a beacon, fostering linguistic diversity and global opportunities in an interconnected world.

## 1 INTRODUCTION

Before learning to walk, a person needs to have a basic understanding of their environment and the ability to process visual and spatial information. Similarly, the same applies to language learning. The mistake is made of teaching grammar before learning to pronounce and understand words. Therefore, teaching phonics for correct pronunciation to a baby facilitates the correct learning and pronunciation of the new language, in our case, English.

The importance of acquiring English as a foreign language in childhood is fundamental in contemporary education. As the world grows more interconnected, English becomes an essential tool for global communication and access to educational and professional opportunities.

Learning a foreign language from an early age has documented cognitive development benefits for children. Within this context, English phonics education emerges as a crucial tool for comprehensive cognitive development, especially in formative years. Phonics, focusing on word sounds, plays a crucial role in teaching English to children. Correct pronunciation, deeply rooted in phonics, is essential for the improvement and enhancement of English pronunciation, particularly in children who naturally grasp foreign languages from an early age. To address this need, a digital platform has been conceived to facilitate English learning through phonics from the age of two onward.

English's lack of direct correspondence between sounds and letters necessitates the teaching of proper pronunciation through phonics. The platform ensures sounds are tailored to be akin to those in Spanish, easing the learning curve for pronunciation.

### 1.1 Theories used in this study

Explores three relevant theories for the study, each contributing valuable insights to the field of early language acquisition:

**Eric Jensen [1]** Jensen applies neuroscience to teaching, emphasizing a positive learning environment and adaptive instruction. Monterrubio specializes in preschool English education, highlighting the importance of adapting teaching styles to cater to various learning methods, be it auditory, visual, or kinesthetics.

These theories collectively provide a comprehensive framework for enhancing language learning effectiveness, considering the complexities of the mind and individual learning styles. The integration of these theories can significantly contribute to improved language acquisition, particularly in early childhood.

**Patricia K. Kuhl's Theory [2]** Kuhl's groundbreaking research focuses on the early language acquisition process in babies. Her emphasis on social and emotional contact as crucial for linguistic learning, coupled with innovative technology use, highlights the importance of a child's intrinsic fascination for the new in establishing a strong language foundation.

**Roger Sperry's Theory [3]** Sperry's holistic educational approach considers not only cognitive but also emotional, social, and physical development. Notably, his theory underscores the significance of early exposure to educational content and understanding diverse learning styles for effective language acquisition, especially in the early developmental stages.

*1.1.1 Does Gamification Work? — A Literature Review of Empirical Studies on Gamification [4]*

The research on gamification has come a long way, but there are still gaps that need to be filled. While it has been shown to have a positive impact, the effects of gamification are highly dependent on the context in which it is used and the users who engage with it. By doing so, we can ensure that gamification is utilized to its fullest potential and that it continues to yield positive results.

These theories collectively provide a comprehensive framework for enhancing language learning effectiveness, considering the complexities of the mind and individual learning styles. The integration of these theories can significantly contribute to improved language acquisition, particularly in early childhood.

## 2  METHODOLOGY

To create educational digital content, we followed a structured process that included the following steps:

- Identify the Target Audience: Our target audience was young children who are learning English phonics in a bilingual education environment.
- Define Educational Objectives: Our educational objectives were clearly defined and included goals such as helping children recognize English letter sounds and pronounce English words correctly.
- Select the Appropriate Digital Management: We chose WordPress, as it was the best fit for our goals.
- Create a Clear and Organized Structure: We developed a clear and organized structure that made it easy for users to navigate and access the content. We used categories, tags, and dropdown menus to organize the content.
- Develop Educational Content: We developed educational content that aligned with our established educational objectives. We incorporated interactive resources such as games, videos, audio, and animations to make learning enjoyable and effective.
- Integrate Tracking and Evaluation Tools: We included tracking and evaluation tools to measure student progress and adapt the content according to their needs.
- Promote Interaction: We promoted interaction among students through collaborative tools such as forums, chats, and wikis.
- Evaluate and Update Content: We periodically evaluated the content to detect potential issues and enhance the learning process. We also updated the content to align it with the needs and demands of the target audience.

Nielsen's 10 Heuristics [5] are a set of usability principles established as guidelines for designing websites and other user interfaces. These principles help create interfaces that are intuitive, efficient, and satisfying for users.

The research illustrated how a well-implemented educational digital web page, like WordPress, could be a transformative force in early-age English education. This was particularly evident when infused with innovative pedagogical approaches and multimedia resources inspired by successful models in the field. We also acknowledged the collaborative role of parents and teachers in shaping a holistic language learning experience for young learners.

## 3  RESULTS

### 3.1  Surveys

https://docs.google.com/forms/d/1X7oO7qBt3ix5AshltXKtKU4jS0ug6wICW8LsADWx4Js/edit?pli=1&pli=1&pli=1#settings

### 3.1.1 Survey questions

1. **What is your opinion on the use of phonics in children's learning?**
   **Result:** Majority in favor (57.9%)

2. **Do you think using codes facilitates English learning for children who are starting to read and write?**
   **Result:** Divided opinion, with a slight leaning towards agreement (50%)

3. **Would creating a digital portfolio not only facilitate learning but also keep necessary materials readily available?**
   **Result:** Majority in favor (57.9%)

4. **If children learn phonics naturally and organically through digital applications, would this improve their pronunciation?**
   **Result:** Majority in favor (72.2%)

5. **Do you think it is necessary to teach vocabulary first before explaining grammar?**
   **Result:** Majority in favor (47.4%)

6. **Do you believe using digital skills would facilitate language learning in 2-year-old children?**
   **Result:** Divided opinion, with a slight leaning towards agreement (57.9%)

### 3.1.2 Conclusions

- Phonics and digital applications are valuable tools for children's language learning.
- There is agreement on the importance of creating digital portfolios for learning.
- Opinions on the order of teaching vocabulary and grammar are divided.
- Using digital skills might be beneficial for language learning in 2-year-olds, but more research is needed.

### 3.1.3 Additional Observations

- The surveys show significant participation, indicating an interest in the topic.
- The responses are varied, reflecting diverse perspectives on children's language learning.
- It is important to consider the limitations of the surveys, such as sample size and methodology.

These results were considered when designing the accessible structure for the project. The diversity of opinions and needs of teachers is considered to ensure that the platform or system developed adequately addresses different perspectives and provides flexible options to adapt to individual preferences. Additionally, it is crucial to offer proper training and support for those teachers who are not familiar with digital applications to make the most of the available tools.

## 3.2 Card Sorting Results [figure 7]

The survey results indicate a strong interest in learning phonetics. Most respondents (75%) are interested in learning about vowels and important sounds. Additionally, a significant portion (62%) are interested in learning phonetics in general.

**However, a gap exists between interest and current knowledge of phonetics.** Only half of the respondents (50%) are familiar with phonetics in teenagers and the use of microphones for pronunciation practice.

**Respondents also expressed a strong preference for personalized and adaptable teaching.** The majority (75%) believe that phonetics teaching should be adapted to the needs and learning pace of each student. Additionally, a considerable portion (37%) are interested in workshops guiding the teaching of phonetics at early ages.

**Regarding content, respondents showed a stronger preference for multimedia resources and interactive activities.** The most popular options were related songs (75%), levelled readings (25%), interactive activities (25%), and related gamification (25%).

Overall, the survey indicates a high demand for phonetics teaching resources and methods that are personalized, engaging, and adaptable to different ages and learning levels.

**Recommendations:**

- Develop a phonetics teaching program that is personalized, adaptable, and engaging for students of different ages and learning levels.
- Include a variety of multimedia resources and interactive activities in the program, such as songs, levelled readings, interactive activities, and gamification.
- Offer workshops and training to teachers and parents on teaching phonetics at early ages.
- Create specific content for 2-year-olds, such as educational songs and videos.
- Research and develop phonetics teaching methods that are effective for students of different ages and learning styles.

This study provides valuable insights for developing phonetics teaching programs that are effective and meet the needs of students.

## 4  WEB DESIGN

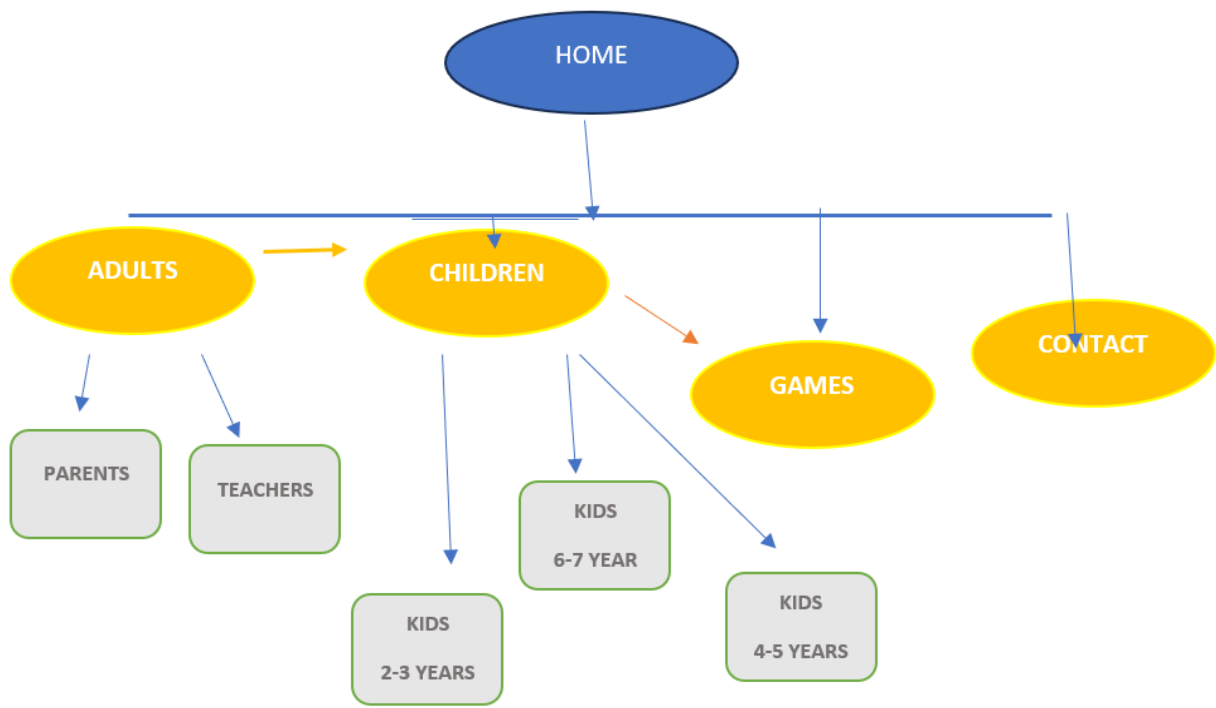

*Figure 1. A diagram illustrating the interaction in the Web.*

## 5  DESCRIPTIVE CONTENT CARDS

The following figures were taken from the following web page related to this article.

https://adelaj.sg-host.com

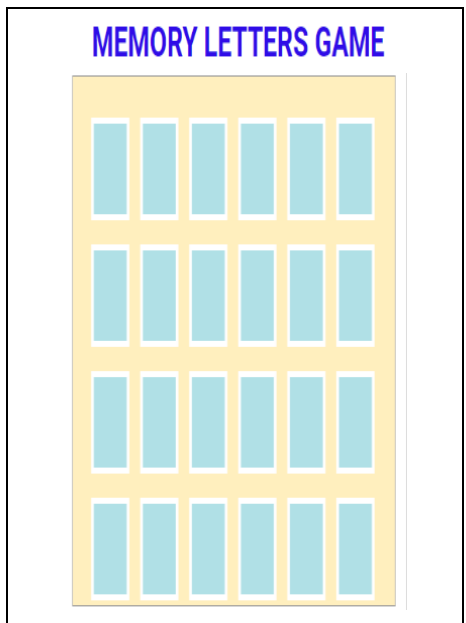

**Description:** This is a memory game that reviews the letters of the English alphabet. It is an application created with HTML5, CSS, and JavaScript. All of the application's code was inserted into additional CSS.

**Learning Objective:** The learning objective of this game is to help students gain confidence and familiarity with reading and writing. For older students, it can also be used as a reinforcement activity for memorizing the letters of the alphabet.

**Content:** The game reviews the letters of the English alphabet.

**Teaching Methodology:** The game is an interactive activity that uses competition to help students develop interest in learning. It also helps them gain confidence and independence, which can help them when learning new topics.

**Prerequisites:** Students should know the sound of each letter of the alphabet before playing the game. This will help them to repeat what they have learned each time they play.

**Additional Information:** The game is designed for students of all ages. A web browser.The game is free to play. The game can be played on any device with

*Figure 1. Memory game for learning the English alphabet.*

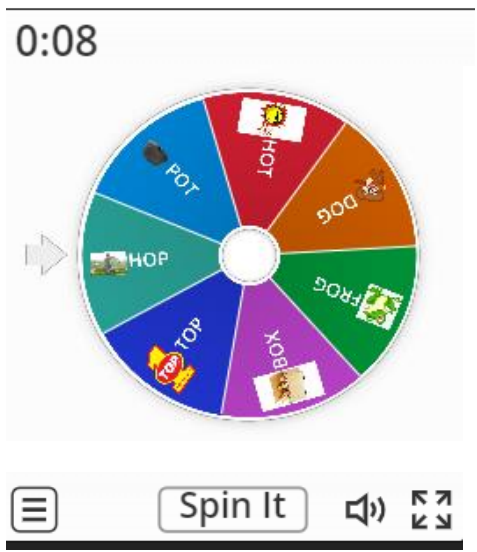

**Description:**

This is a roulette game that reviews the short sounds of the vowels a, e, i, o, and u. There are 5 identical containers.

**Learning Objective:** The learning objective of this game is to help students gain confidence and familiarity with reading and writing. For older students, it can also be used as a reinforcement activity for memorizing the vowels.

**Content:** The game reviews the vowels a, e, i, o, and u.

**Teaching Methodology**: The game is an interactive activity that uses competition to help students develop interest in learning. It also helps them gain confidence and independence, which can help them when learning new topics. Students can guess which word will come up when they play the game.

**Prerequisites:** There are no prerequisites for this game**.**

**Additional Information:** The game is designed for students of all ages. The game can be played on any device with a web browser. The game is free to play.

*Figure 2. Short vowels roulette game.*

A widely used application in education, especially for young children, is very interactive and with the new updates it has been possible to upload them to WordPress.

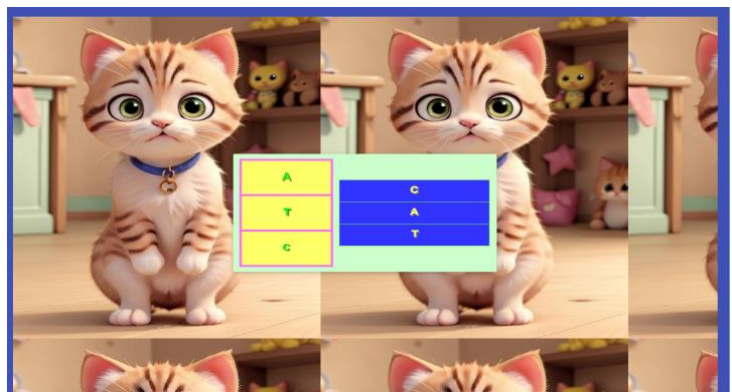

*Figure 3. Educational application for children.*

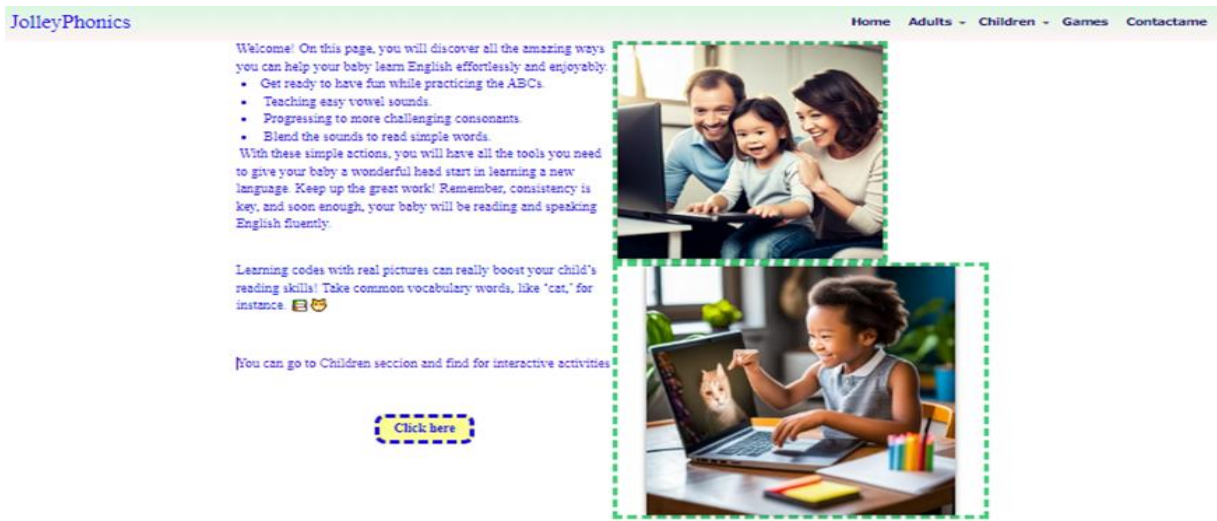

*Figure 4. Parent's section of the platform.*

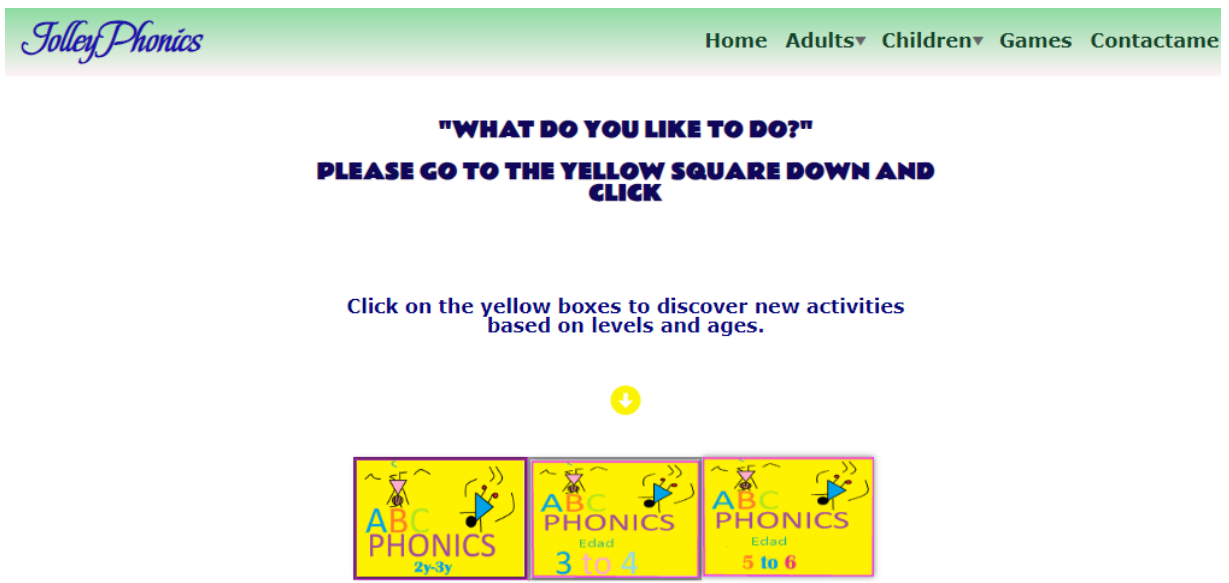

*Figure 5. Children's section of the platform.*

## 6 CONCLUSIONS

Several theories suggest that younger children can learn a language naturally and perfectly. Therefore, it is essential to use digital technology to provide interactive and engaging English language learning opportunities to children. To this end, a user-friendly WordPress page has been created, offering playful and interactive activities designed to capture children's attention and create a friendly learning environment.

Based on the results of card sorting studies, a help section for teachers has been included to motivate them and provide confidence when teaching English to young children. The platform also can contain a progress evaluation by observing the skills the child develops through various activities, this evaluation will be given in future updates.

The project's structure has been designed based on the diverse opinions and needs of teachers, ensuring that the platform addresses different perspectives and provides flexible options to adapt to individual preferences. The selection of educational material on the website prioritizes usability and accessibility, using visual resources and codes to make the process enjoyable for children.

Special attention has been given to vocabulary learning, particularly for Spanish-speaking children, who may face challenges when learning English. To overcome these challenges, the website offers resources and activities that make learning a new language a friendlier and more understandable experience.

Visual, interactive, and technological elements like activities and games have been included to encourage the active participation of children, making the vocabulary learning process more dynamic and engaging. The website's intuitive navigation system simplifies access to content and promotes active participation, contributing to effective and satisfying learning of English vocabulary.

## ACKNOWLEDGEMENTS

To my Doctoral School Directors Dr. Pedro Paredes, and Codirector Dra. Virginia Vinuesa.

## WEBPAGE

https://adelaj.sg-host.com

# APPENDIX

Card Sorting Results. (Since this research is for Spanish, it is in Spanish)

- **Vocales: A, E, I, O, U. cortes** (Vowels: A, E, I, O, U. cuts) - 75%
- **Sonidos importantes** (Important sounds) - 75%
- **Aprendizaje de fonética** (Phonetics learning) - 62%
- **Fonética en adolescentes** (Phonetics in adolescents) - 50%
- **Micrófono para la pronunciación** (Microphone for pronunciation) - 50%
- **Adapter la enseñanza de la fonética a las necesidades y ritmo de aprendizaje de cada estudiante** (Adapt the teaching of phonetics to the needs and learning pace of each student) - 37%
- **Lecturas de acuerdo a los niveles de fonética** (Readings according to phonetics levels) - 25%
- **Actividades interactives te gustaría e** (Interactive activities you would like) - 25%
- **Gamificacióm relacionada** (Related gamification) - 25%
- **Creación de un Avatar guía para el viaje de este aprendizaje** (Creation of an Avatar guide for the journey of this learning) - 25%
- **Crear niveles según aprendizaje** (Create levels according to learning) - 25%
- **Organizar formatos de contenido prefieres para el aprendizaje** (Organize content formats you prefer for learning) - 25%
- **Profesores, Padres, y niños de todas las edades** (Teachers, Parents, and children of all ages) - 0%
- **Contenido especial para niños de 2 años** (Special content for children 2 years old) - 25%
- **Inglés para niños desde 2 años** (English for children from 2 years old) - 25%
- **Canciones relacionadas** (Related songs) - 75%
- **Sonidos de apoyo** (Support sounds) - 25%
- **Workshop guia la enseñanza de la fonética en edades tempranas** (Workshop guide for the teaching of phonetics in early ages) - 37%
- **Contenido relacionado con la fonética inglesa y la educación bilingüe** (Content related to English phonetics and bilingual education) - 25%

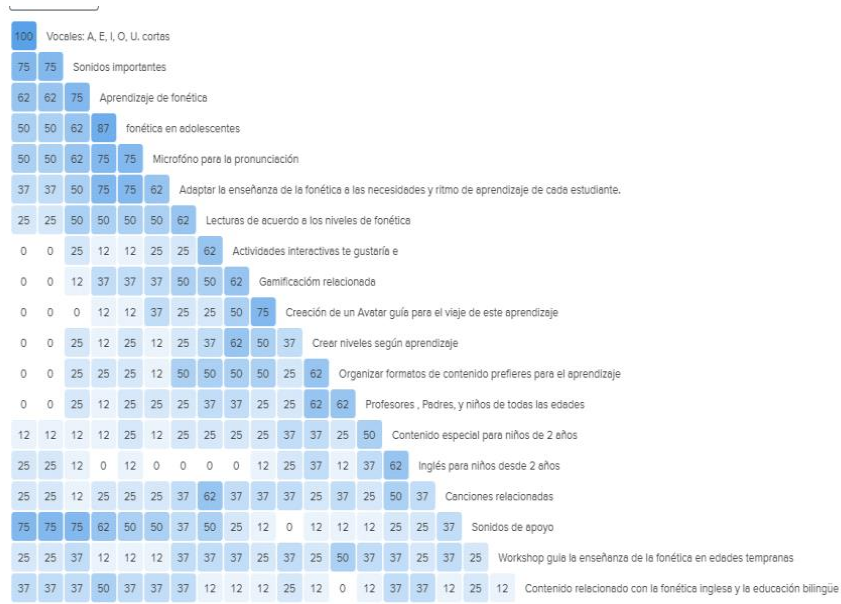

*Figure 6. The interaction between adults and children*

This table shows the results of a card sorting activity where 28 parents were asked to categorize different resources for learning English. The percentages indicate the proportion of parents who chose each option.